\documentclass[prl,twocolumn,nopacs,preprintnumbers,notitlepage,amsmath,amssymb,superscriptaddress,nofootinbib]{revtex4-1}

\usepackage{amssymb}
\usepackage{amsmath}
\usepackage{color}
\usepackage[pdftex]{graphicx}

\usepackage{empheq}

\usepackage{mathtools}

\usepackage{makecell}

\usepackage{enumerate}
\usepackage{mathtools}
\usepackage{stmaryrd}

\usepackage{enumitem}

\usepackage{textcomp}
\usepackage{gensymb}

\newcommand{\ex}[1]{\mathrm{e}^{#1}}

\newcommand{\dd}[0]{\mathrm{d}}
\newcommand{\ii}[0]{\mathrm{i}}
\newcommand{\kk}[0]{\boldsymbol{k}}

\newcommand{\zz}[0]{\mathbf{0}}

\newcommand{\ee}[0]{\boldsymbol{e}}
\newcommand{\rr}[0]{\boldsymbol{r}}

\newcommand{\xx}[0]{\boldsymbol{x}}

\newcommand{\yy}[0]{\boldsymbol{y}}

\newcommand{\sgn}[0]{\mathrm{sgn}}

\newcommand{\XX}[0]{\boldsymbol{x}}
\newcommand{\YY}[0]{\boldsymbol{y}}

\definecolor{darkblue}{rgb}{0,0,0.6}
\definecolor{darkred}{rgb}{0.6,0,0}
\usepackage[colorlinks=true,urlcolor=darkblue,citecolor=darkblue,linkcolor=darkred]{hyperref}

\begin{document}

\title{Exact time dependence of the cumulants of a tracer position in a dense lattice gas}

\author{Alexis Poncet}
\affiliation{Univ. Lyon, ENS de Lyon, Univ. Claude Bernard,
CNRS, Laboratoire de Physique, 69342 Lyon, France}
\author{Aur\'elien Grabsch}
\affiliation{Sorbonne Universit\'e, CNRS, Laboratoire de Physique Th\'eorique de la Mati\`ere Condens\'ee (LPTMC), 4 Place Jussieu, 75005 Paris, France}

\author{Olivier B\'enichou}
\affiliation{Sorbonne Universit\'e, CNRS, Laboratoire de Physique Th\'eorique de la Mati\`ere Condens\'ee (LPTMC), 4 Place Jussieu, 75005 Paris, France}

\author{Pierre Illien}
\affiliation{Sorbonne Universit\'e, CNRS, Physicochimie des Electrolytes et Nanosyst\`emes Interfaciaux (PHENIX), 4 Place Jussieu, 75005 Paris, France}

\date{\today}

\begin{abstract}
We develop a general method to calculate the exact time dependence of the cumulants of the position of a tracer particle in a dense lattice gas of hardcore particles.  More precisely, we calculate the cumulant generating function associated with the position of a tagged particle at arbitrary time, and at leading order in the density of vacancies on the lattice. In particular, our approach gives access to the short-time dynamics of the cumulants of the tracer position -- a regime in which few results are known.  The generality of our approach is demonstrated by showing that it goes beyond the case of a symmetric 1D random walk, and covers the important situations of (i) a biased tracer; (ii) comb-like structures; and (iii) $d$-dimensional situations.
\end{abstract}

\maketitle

%\tableofcontents

\emph{Introduction.---} Understanding and characterising tracer diffusion in crowded environments is central in numerous biological and physical contexts. In living systems, the interplay between the diffusion of tracer particles (fuelled by thermal fluctuations, active processes, or chemical reactions) and complex environments (which generally hinder their motion) controls many biological processes~\cite{Hofling2013}. Quantifying tracer diffusion can also be used as a mean to probe the mechanical and rheological properties of different systems, such as colloidal suspensions or complex fluids, through passive and active microrheology~\cite{{Wilson2011},{Squires2010},Puertas2014a}. 

These examples, in which the statistical properties of tracer particles are controlled by the interactions with their environment, are the motivation for a whole field of theoretical research. Among the different routes that were employed to characterise the statistics of diffusing particles in crowded environments, lattice gases of hardcore particles that jump at \emph{exponentially distributed times} (often referred to as {exclusion processes}) have been the subject of many studies, and have become central models of statistical mechanics~\cite{Mallick2011,Chou2011}. In particular, such models were widely employed to compute the diffusion coefficient of a tracer particle. In dimension 2 or greater, different mean-field-like approximations were designed  to estimate the diffusion coefficient of the tracer as the function of the density of the bath~\cite{Nakazato1980,Tahir-Kheli1983a,vanBeijeren1985a}. In the case of one-dimensional systems, one can mention recent achievements which resulted in the derivation of exact results concerning tracer properties, including the calculation of its large deviations~\cite{Imamura2017, Krapivsky2015, Krapivsky2014, Hegde2014}, and of bath-tracers correlations~\cite{Poncet2021,{Grabsch2021}}.

\begin{figure}
	\centering
	\includegraphics[width=\columnwidth]{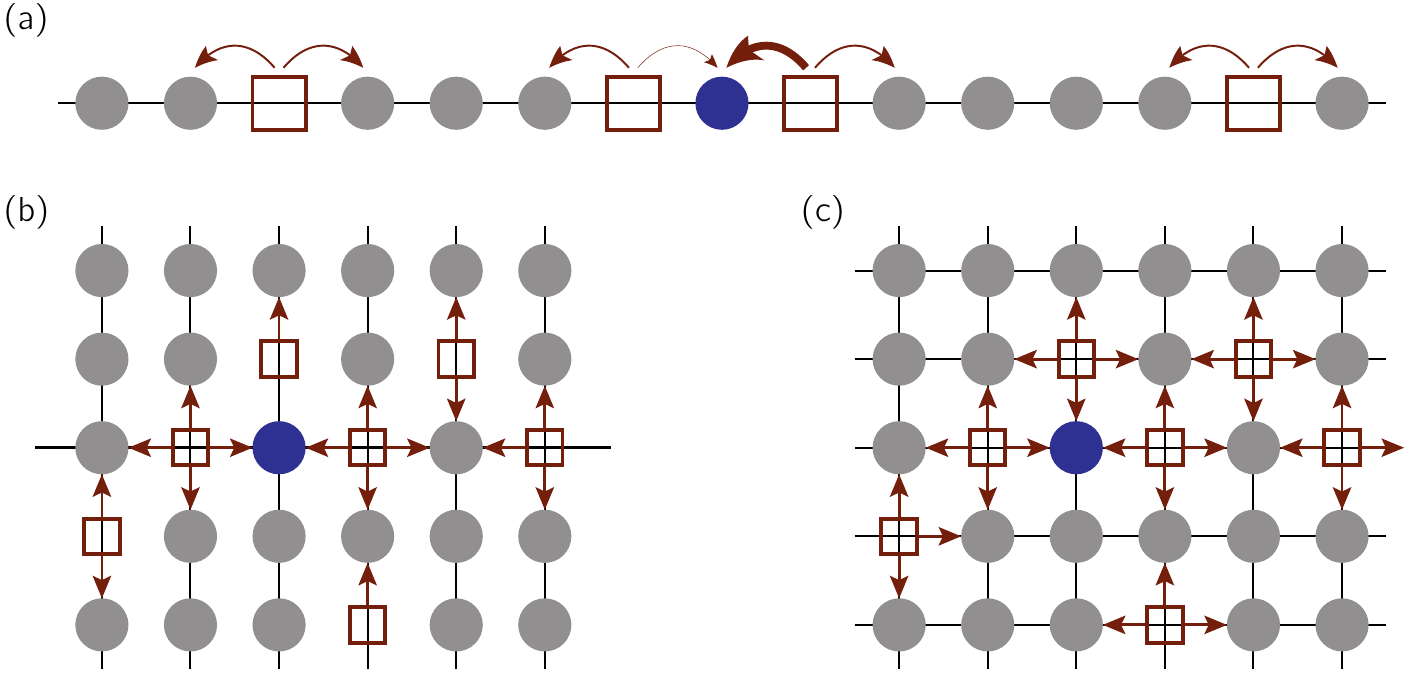}
	\caption{For each of the considered geometries (1D (a), comb (b), 2D (c)), the {continuous-time} random walks of the particles are mirrored by the random walks of the vacancies (brown squares). The latter perform continuous-time nearest-neighbor symmetric random walks.}
	\label{fig:1tp_vacs}
\end{figure}

However, these results, whether exact or approximate, are generally valid only in the long-time limit, because their derivation relies on hydrodynamic limits or large deviations approaches~\cite{Imamura2017, Krapivsky2015, Krapivsky2014, Hegde2014,Poncet2021,{Grabsch2021}}. A notable exception is provided in the dense limit by the approach by Brummelhuis and Hilhorst ~\cite{Brummelhuis1989a,Brummelhuis1988}, later extended to the case of a biased tracer  \cite{{Benichou2002a}, Illien2013a,Benichou2013c,Illien2014}. However, we stress that this approach is intrinsically \emph{discrete in time}. Even though the real continuous-time description of exclusion processes, where particles jump at exponential times as defined above, is retrieved in the long-time limit, this approach fails to predict the dynamics of the tracer at short and intermediate times.
So far, the only available results at arbitrary time concern the first cumulants in the low-density regime with immobile bath particles~\cite{Leitmann2017,Leitmann2013}, the high-density regime for a symmetric tracer in 1D~\cite{Poncet2021}, or the 1D situation at arbitrary density, but under a formulation that does not allow the derivation of fully explicit results~\cite{Imamura2021}.  Finally, a general quantitative description of the dynamics of the tracer for arbitrary time is lacking.

 In this Letter, we fill this gap and calculate the exact and complete time dependence of the cumulants of a tracer particle in a dense lattice gas. We develop a general methodology which covers the important cases of (i) a biased tracer; (ii) comb-like structures; and (iii) $d$-dimensional situations.  These results fully quantify the dynamics of tracer particles in exclusion processes, which are paradigmatic models of  statistical mechanics.

\emph{Model and outline of the calculations.---}
We consider a lattice populated by particles at a density $\rho$ between 0 and 1, which are initially positioned uniformly at random on the lattice, with the restriction that there can only be one particle per site. 
We adopt the  usual dynamics of exclusion processes, which evolve in continuous-time, and we assume that each particle has an exponential clock of time constant $\tau=1$. When the clocks ticks, each particle chooses to jump on one of its $z$ neighboring sites with probability $1/z$. If the arrival site is empty, the jump is done. Otherwise, if the arrival site is occupied, the jump is canceled. Note that, in 1D, this process corresponds to the celebrated Symmetric Exclusion Process or SEP~\cite{Mallick2011,Chou2011}. 

 The tagged particle (TP) is initially at the origin $\XX(0) = 0$ and we study its displacement with time $\XX(t)=(x_1(t),\dots,x_d(t))$. We define the cumulant-generating function (CGF) $\psi(\kk,t) \equiv \ln \langle e^{i \kk \cdot \XX(t)}\rangle$.
We will consider the cumulants of the position projected onto one direction of the lattice, say direction 1 ($x_1(t) = \xx(t)\cdot\ee_1$):
\begin{equation}
\label{ }
\kappa_n(t) = \frac{1}{\ii^n} \left( \frac{\partial^n \psi(\kk,t)}{\partial {k_1}^n}\right)_{\kk=\zz}.\end{equation}
Our goal here is the determination of the cumulant-generating function $\psi(\kk,t)$ and the cumulants $\kappa_n(t)$ in the high-density limit $\rho\to 1$, and for arbitrary time $t$. We will define their rescaled high-density limit as $\bar \kappa_n = \lim_{\rho\to1} \kappa_n/(1-\rho)$, where $\rho_0=1-\rho$ is the density of vacancies on the lattice.

\emph{From a single vacancy to the dense regime.---} Relying on the derivation that was originally proposed in a discrete-time description~\cite{Brummelhuis1988,Brummelhuis1989a}, we start by considering a system of finite size $N$ in which all the sites are occupied except $M$ of them (Fig. \ref{fig:1tp_vacs}). We call these empty sites \emph{vacancies}, and their fraction is denoted by $\rho_0 = M/N = 1-\rho$. Now, in the high-density limit ($\rho_0 = M/N\to 0$), we note that
the vacancies perform independent random walks and interact independently with the TP.
We neglect events of order $\mathcal{O}(\rho_0^2)$ in which two vacancies interact with each other, compared to events of order $\mathcal{O}(\rho_0)$ in which one vacancy interacts with the TP. This gives exact results at linear order in the density of vacancies $\rho_0$. We call $p_1(\XX|\YY,t)$ the probability that, in a system with a single vacancy initially at $\YY$, the TP has reached site $\XX$ at time $t$ knowing that it started from the origin. In Fourier space, the probability to find the tracer at a given location given that the vacancies were initially at positions $\yy_1,\dots,\yy_M$ can be written as a product of single-vacancy propagators $p_1$ (see Section I in Supplemental Material (SM)~\cite{SM}). Averaging over the initial positions of the vacancies and taking the thermodynamic limit of $M,N\to\infty$ with fixed $\rho_0$, the cumulant-generating function reads 
\begin{equation} \label{eq:1tp_expr_psi}
\lim_{\rho_0\to 0} \frac{\psi(\kk,t)}{\rho_0} = \sum_{\YY\neq 0} \left(\tilde p_1(\kk|\YY,t)-1\right),
\end{equation}
where we use the following convention for Fourier transforms $\tilde f(\kk) = \sum_{\XX} \ex{\ii \kk\cdot\XX}f(\XX)$. Let us emphasize the meaning of Eq. \eqref{eq:1tp_expr_psi}: the full probability law of a TP at high density is encoded in a much simpler quantity, namely the propagator of the tracer in a system where there is only one vacancy. This expression is the continuous-time counterpart of the discrete-time approach~\cite{Brummelhuis1988, Brummelhuis1989a}. 

Using standard techniques from the theory of random walks on lattices \cite{Hughes1995}, the single-vacancy propagators $p_1(\xx|\yy,t)$ can be expressed in terms of first-passage time densities associated with the random walk performed by a vacancy on the considered lattice, namely $f(\zz|\yy,t)$, the probability for the vacancy to reach the origin for the first time at time $t$ knowing that it started from site $\yy$, and $f^*(\zz|\ee_\mu|\yy,t)$, the same quantity but conditioned on the fact that the vacancy was at site $\ee_\mu$ before its last jump. The relation between these quantities is obtained by counting the interactions between the vacancy and the tracer up to time $t$.

For clarity we consider separately: (i) the situation where the lattice is \emph{tree-like}, i.e. the situation where there is a single minimum-length path linking two arbitrary sites of the lattice (this will cover the case of one-dimensional and comb-like lattices); (ii) the situation where the lattice is \emph{looped}, i.e. the situation where there is more than one minimum-length path linking two arbitrary sites of the lattice (this will cover the case of lattices of dimension 2 and higher).

\emph{Tree-like lattices.---}  
We first consider tree-like lattices as shown in Figs. \ref{fig:1tp_vacs}(a) and \ref{fig:1tp_vacs}(b). We show in SM (Section II of~\cite{SM}) that, on these geometries, the single-vacancy propagator is simply related to the FPT densities through the relation
\begin{equation}
\hat{\tilde p}_1(k|\YY, u) = \frac{1}{u}\left[
1+ \left(e^{i\mu k} -1\right) \frac{1-\hat f_{-\mu}(u)}{1-\hat f_1(u)\hat f_{-1}(u)}\hat f(\zz|\YY,u) \right],
\label{eq:1tp_expr_pZ}
\end{equation}
where we introduce the shorthand notation $ \hat{f}_\nu(u)=\hat f(\zz|\ee_\nu,u)$. One can use this expression into Eq.~\eqref{eq:1tp_expr_psi} to obtain the cumulant-generating function at high density in terms of first-passage quantities of a single vacancy. The last step consists in studying the random walk of a single vacancy to compute $\hat f(\zz|\yy,u)$.

\begin{figure}
\begin{center}
\includegraphics[width=\columnwidth]{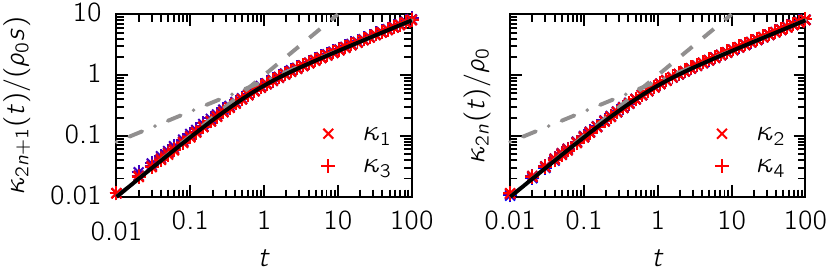}
\caption{Time dependence of the odd (left) and even (right) cumulants of a TP in 1D ($\rho_0 =0.02$) for different values of the bias (from blue to red: $s=0$, $0.2$, $0.5$, $0.8$, $1$).  Symbols are the results from numerical simulations (see Section IX of~\cite{SM} for details). The black lines are the predictions from Eqs. \eqref{eq:1tp_resCum_even} and \eqref{eq:1tp_resCum_odd}, the gray lines are the asymptotic regimes at short and large times.}
\label{fig:sep}
\end{center}
\end{figure}

We first apply this formalism to the case of a 1D lattice (Fig. \ref{fig:1tp_vacs}(a)). We consider the general situation of a biased tracer which jumps with probability $p_+$ to the right and $p_{-}$ to the left. A vacancy then performs a nearest-neighbor random walk, which is symmetric far away from the tracer and perturbed in its vicinity. Considering a vacancy starting from site $y$ and partitioning over the instant of first visit to site $\mu=\sgn(y)$, it is straightforward to show that $\hat f_y(u) = \hat f^\mathrm{UB}_{|y|-1}(u) \hat f_\mu(u) $~\cite{SM}, where we introduced the following notation for the Laplace transform: $\hat{\varphi}(u) = \int_0^\infty \ex{-ut}\varphi(t)\dd t$, the shorthand notation $f_y(t) = f(0|y,t)$, and where the superscript UB denotes the FPT density of the vacancy to the origin when the tracer is unbiased. We finally get (Section III in~\cite{SM}):
\begin{equation} \label{eq:1tp_res_fZ}
\hat f_y(u) = \frac{1+\mu s}{1+\mu s\alpha} \alpha^{|y|},
\end{equation}
where $\alpha = 1 + u - \sqrt{u(2+u)}$ and where $s = p_+ - p_-$ is the bias. Inserting the first-passage quantities computed in Eq.~\eqref{eq:1tp_res_fZ} into the expression of the propagator with a single vacancy [Eq.~\eqref{eq:1tp_expr_pZ}], and then back into the expression of the cumulant-generating function [Eq.~\eqref{eq:1tp_expr_psi}], we obtain, after Laplace inversion:
\begin{equation}
\lim_{\rho_0\to 0} \frac{\psi(k, t)}{\rho_0}
= te^{-t}[\text{I}_0(t) + \text{I}_1(t)](\cos k - 1 +is\sin k), 
\label{cgf_1D}
\end{equation}
where $\text{I}_0$ and $\text{I}_1$ are modified Bessel functions of the first kind~\cite{AbramovitzI.1972}. In the unbiased case $s=0$, we retrieve previous results for a symmetric tracer in the SEP~\cite{Poncet2021}. The first implication is that we have the full time-dependence of the even and odd cumulants,
\begin{eqnarray} 
 {\bar \kappa_{2n}(t)}
&= & te^{-t}[\text{I}_0(t) + \text{I}_1(t)],\label{eq:1tp_resCum_even} \\
 {\bar \kappa_{2n+1}(t)}
&= & ste^{-t}[\text{I}_0(t) + \text{I}_1(t)].\label{eq:1tp_resCum_odd} 
\end{eqnarray}
At short time, we find that the cumulants obey $\bar \kappa_{2n}(t) \sim t$ and  $\bar \kappa_{2n+1}(t) \sim st$.
This means in particular that the fluctuations of the tracer are diffusive, and that the displacement of a biased TP $\kappa_1$ is ballistic. At large time, we retrieve the known expressions~\cite{Arratia1983,Illien2013a,Imamura2017}: $\bar \kappa_{2n}(t) \sim  \sqrt{{2t/\pi}}$ and $\bar \kappa_{2n+1}(t) \sim s \sqrt{{2t/\pi}}$. At all times, the results from Eqs. \eqref{eq:1tp_resCum_even} and \eqref{eq:1tp_resCum_odd} are in perfect agreement with numerical simulations and shown on Fig. \ref{fig:sep}.

\begin{figure}
\begin{center}
\includegraphics[width=\columnwidth]{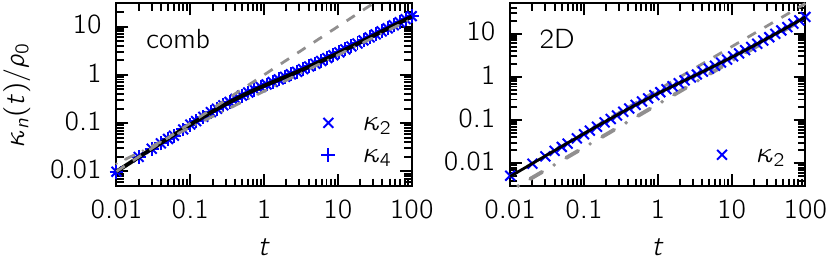}
\caption{Cumulants of the position of a tracer constrained to move on the backbone of a comb lattice (left) and on a 2D lattice (right). Vacancy density $\rho_0 =0.01 $. The symbols correspond to the results from numerical simulations, the solid line is obtained from the inversion of the expression in Laplace domain [Eq. \eqref{cgf_comb3} and \eqref{2D_kappa2_Laplace}] using the Stehfest algorithm. The short-time (dashed line) and long-time (dash-dotted line) asymptotics are given in the text.}
\label{fig:comb_and_2D}
\end{center}
\end{figure}

We further illustrate the generality of our method by considering the important case of a comb lattice, a lattice made of a line, called the backbone, on which other lines, called the teeth, are connected (Fig. \ref{fig:1tp_vacs}(b)). This structure has been widely used to describe diffusion on percolation clusters~\cite{Ben-avraham}. From now on and for simplicity, we restrict ourselves to the case of a symmetric tracer constrained to move on the backbone of the lattice. Relying on the same methodology as before, the density of first-passage time to the origin of a vacancy starting from site $(y_1,y_2)$ reads (Section IV of~\cite{SM})
\begin{equation}
\hat{f}(0,0|y_1,y_2;u)=
\begin{cases}
   \hat{f}_1 (\hat{f}_\parallel)^{|y_1|-1} \hat{f}_\perp\alpha^{|y_2|-1} & \text{if $y_2\neq0$}, \\
   \hat{f}_1     (\hat{f}_\parallel)^{|y_1|-1} & \text{if $y_2= 0$}.
\end{cases}
\label{fpt_comb}
\end{equation}
where we introduced the shorthand notations $\hat{f}_{\mu}=\hat{f}(0,0|\mu,0,u)$, $\hat{f}_{\parallel}= \hat{f}(1,0|2,0,u) $ and $ \hat{f}_{\perp} = \hat{f}(1,0|1,1,u) $, which can all be easily expressed in terms of $\alpha$ (Section IV in~\cite{SM}).
Introducing Eq.~\eqref{fpt_comb} into Eq.~\eqref{eq:1tp_expr_pZ}, we finally obtain
\begin{align}
  \label{cgf_comb3}
&\lim_{\rho_0 \to 0} \frac{\psi(k,u)}{\rho_0} =
 \hat K(u) (\cos k -1), \\
&\text{with~}\hat K(u) = (2-\alpha)(\alpha^2-\alpha+2)/\{u(\alpha-1) \nonumber\\
&\times [u(2-\alpha)+\beta-4\alpha+6][u(\alpha-2)+\beta+2\alpha-2]\}, \nonumber
\end{align}
where $\beta = \sqrt{[(2+u)\alpha-2u-2][(3+u)\alpha-2u-4]}$.
While odd cumulants are null (for symmetry reasons, and as can be seen from Eq. \eqref{cgf_comb3}), all the even cumulants are equal, and given by $\hat{\bar \kappa}_\text{even}(u) = \hat K(u)$.
We deduce, after Laplace inversion, the short-time and long-time expansions: $K(t) \underset{t\to 0}{\sim} t $ and $K(t) \underset{t \to \infty}{\sim} \frac{2^{3/4}}{3\Gamma(3/4)}t^{3/4} $.
Note that the long-time limit in the case of a symmetric tracer corresponds to the result we derived in discrete time~\cite{Benichou2015}. For arbitrary time, we invert the cumulants numerically using the Stehfest algorithm. Numerical simulations are in perfect agreement with our analytical results (Fig. \ref{fig:comb_and_2D}). Note that it is known that the two limits $t\to\infty$ and $\rho_0 \to 0$ do not commute, which mirrors the existence of a subtle ultimate diffusive regime~\cite{Benichou2015}, that we do not intend to describe here.

\emph{Looped lattices.---} We finally consider the key situation of $d$-dimensional lattices. Note that the geometry can be general, each of the spatial directions of the lattice being either be infinite or finite with periodic boundary conditions, in such a way that the lattice remains translation-invariant.  The CGF of the position of a symmetric tracer now reads (Section V in~\cite{SM})
\begin{equation}
 \lim_{\rho_0 \to 0} \frac{\hat{\psi}(\kk,u)}{\rho_0} = -\sum_{j=1}^d \hat{\Delta}(\kk|\ee_j,u) \hat{f}'_j(u),
    \label{cgf_Delta_fp}
\end{equation}
where we defined $ f'_\nu(t) = \sum_{\YY\neq\zz} f^*(\zz|\ee_\nu|\YY;t)$ and
\begin{align}
 &   \hat{\Delta}(\kk|\ee_j,u)=\frac{2(1-\cos q_j)}{u}-\frac{1}{u} \sum_{\mu,\nu} 
\left[1-\ex{-\ii \kk \cdot \ee_\nu}\right] \nonumber\\
&\times \left\{[{\bf I} -{\bf T}]^{-1}\right\}_{\nu,\mu}  \ex{\ii {\kk} \cdot \ee_\mu}\sum_{\epsilon=\pm1} \ex{-\epsilon \ii q_j} \hat{f}^*(\zz|\ee_\mu|\epsilon \ee_j,u),
    \label{Delta_fstar}
\end{align}
where $\mathbf{I}$ is the identity of size $2d$ and the matrix ${\rm {\bf T}}$ has the entries ${\rm {\bf T}}_{\mu,\nu}= \ex{\ii \kk \cdot {\bf e_{\nu}}}  \hat{f}^*({\bf 0}|\ee_\nu|-\ee_\mu ; u) $ .

The final step of the calculation consists in determining the conditional FPT $f^*$ in terms of well-known quantities, namely the propagators associated to a discrete-time random walk on a lattice. The starting point of this calculation is the following relation, which consists in partitioning the random walk performed by the vacancy over the time of first visits to the origin:
\begin{align}
  & \int_0^t \dd t_0  \frac{ \chi(t_0)}{2d} p(\ee_\mu|\yy,t-t_0) = \int_0^t \dd t_0 f^*(\zz|\ee_\mu|\yy,t_0)\Psi(t-t_0)\nonumber\\
   &+\int_0^t\dd t_0 \int_0^{t-t_0}\dd t_1\; \frac{1}{2d} \chi(t-t_0-t_1) f(\zz|\yy,t_0) p(\ee_\mu|\zz,t_1),
    \label{partition_fstar}
\end{align}
where $\Psi(t) = 1-\int_0^t \dd t'\; \chi(t')$ is the probability that the walker did not move during a time $t$.  It is then straighforward to express the conditional FPTs $f^*$ in terms of the continuous-time occupation probabilities $p(\rr|\rr_0,t)$ (probability to find a vacancy at site $\rr$ at time $t$ knowing that it started from site $\rr_0$). Finally, relying on the relation between the propagators $p$ and their discrete-time counterpart $P_{\rr}^{(n)}$ (probability to find the walker at site $\rr$ after $n$ steps knowing that it started from the origin) \cite{Hughes1995}, we get the relations (Section VI in~\cite{SM})
\begin{eqnarray} \label{fhat_Phat}
  \hat{f}^*(\zz|\ee_\mu|\yy,u) &=&\frac{\hat{\chi}}{2d}  \left[  \hat{P}_{\ee_\mu-\yy}(\hat{\chi})-\frac{\hat{P}_{\yy}(\hat{\chi})\hat{P}_{\ee_\mu}(\hat{\chi})}{\hat{P}_{\zz}(\hat{\chi}(u))} \right], \\
      \hat{f}'_\mu(u)& =& \frac{1}{2d} \frac{\hat{\chi}}{1-\hat{\chi}} \left[1- \frac{\hat{P}_{\ee_\mu}(\hat{\chi})}{\hat{P}_{\zz}(\hat{\chi})} \right],
      \label{fprimehat_Phat}
\end{eqnarray}
where  $\hat{P}_{\rr}(\xi) = \sum_{n=0}^\infty P^{(n)}_{\rr} \xi^n$ is the generating function associated to the discrete-time propagator $P^{(n)}_{\rr}$.

In summary, the cumulant generating function of the tracer position is fully determined in terms of the generating functions $\hat{P}$ associated to a discrete-time random walk on the considered lattice. Indeed, the expression of the CGF given in Eq. \eqref{cgf_Delta_fp} simply involve $f'$ and $\Delta$. The former is related to the generating functions $\hat{P}$ through Eq. \eqref{fprimehat_Phat}. The latter is related to the conditional first-passage densities $f^*(\zz|\ee_\mu|\YY,t)$ through Eq. \eqref{Delta_fstar}, which are themselves related to the generating functions $\hat{P}$ through Eq. \eqref{fhat_Phat}. This result holds for any translation-invariant lattice, in arbitrary space dimension.

Applying this procedure to the case of a 2D lattice, and making use of the symmetries of the system, one can show that the CGF is expressed in terms of only three first-passage time densities, namely $f^*(\zz|\ee_1|\ee_1,u)$, $f^*(\zz|\ee_1|-\ee_1,u)$ and $f^*(\zz|\ee_1|\ee_2,u)$. These quantities are themselves expressed in terms of the discrete-time propagators  $\hat{P}$ (Section VII in~\cite{SM}), which are simply given by Fourier integrals. As an example, we get the following expression of the second cumulant
\begin{equation}
    \lim_{\rho_0 \to 0} \frac{\hat\kappa_2(u)}{\rho_0} = \frac{1}{2u} \frac{\hat\chi(u)}{1-\hat\chi(u)} \frac{2-\hat\chi(u)g(\hat\chi(u))}{2+\hat\chi(u)g(\hat\chi(u))}
    \label{2D_kappa2_Laplace}
\end{equation}
where $g(\xi)=[\hat P_{\zz} (\xi) - P_{2 \ee_1} (\xi) ]/2$ and is given by the integral quantity~\cite{Brummelhuis1988,Hughes1995}
\begin{equation}
    g(\xi) = \frac{1}{(2\pi)^2} \int_{-\pi}^\pi \dd q_1 \int_{-\pi}^\pi \dd q_2 \; \frac{\sin^2 q_1}{1- \frac{\xi}{2}(\cos q_1 + \cos q_2)}.
    \label{def_g}
\end{equation}
An explicit expression of $g(\xi)$ in terms of elliptic integrals, as well as its asymptotic expansions when $\xi\to0$ and $\xi\to 1$, is given in~\cite{SM} (Section VIII). This  yields in particular the following asymptotics for the second cumulant in Laplace domain : $ \hat \kappa_2(t) \underset{t\to 0}{\sim} {t}/{2}$ and $\hat \kappa_2(t) \underset{t\to \infty}{\sim} {t}/[2(\pi-1)]$.
The expression given in Eq. \eqref{2D_kappa2_Laplace} can be inverted back numerically into the time domain. The output of this inversion, together with numerical simulations and the short-time and long-time asymptotics, are represented on Fig. \ref{fig:comb_and_2D}. The fluctuations of the tracer position go from one diffusive regime to another, and one observes that the long-time diffusion coefficient is approximately half the short-time diffusion coefficient.

\emph{Conclusion.---} In this Letter, we presented a new methodology to compute the full and exact time dependence of the position of a tracer particle in a dense lattice gas. We demonstrated the generality of this method by considering different geometries (1D, comb-like, $d$-dimensional), and obtaining fully explicit expressions (either in Laplace domain or in time domain) for the cumulants of the tracer position. These results unveil the transient time regimes that precede the long-time asymptotics which are usually the only results that can be obtained from the  standard approaches, such as hydrodynamic limits, large deviations, or discrete-time vacancy mediated diffusion. Although the method presented here holds in the dense limit, our results constitute a significant step in the description of the full time dynamics of tracer particles in exclusion processes.

\bibliographystyle{apsrev4-1}
\bibliography{/Users/pierreillien/work/docs/library.bib}

\end{document}

\title{Exact time dependence of the cumulants of a tracer in a dense lattice gas \\ \textit{Supplemental Material}}

\author{Alexis Poncet}
\affiliation{Univ. Lyon, ENS de Lyon, Univ. Claude Bernard,
CNRS, Laboratoire de Physique, 69342 Lyon, France}
\author{Aur\'elien Grabsch}
\affiliation{Sorbonne Universit\'e, CNRS, Laboratoire de Physique Th\'eorique de la Mati\`ere Condens\'ee (LPTMC), 4 Place Jussieu, 75005 Paris, France}

\author{Olivier B\'enichou}
\affiliation{Sorbonne Universit\'e, CNRS, Laboratoire de Physique Th\'eorique de la Mati\`ere Condens\'ee (LPTMC), 4 Place Jussieu, 75005 Paris, France}

\author{Pierre Illien}
\affiliation{Sorbonne Universit\'e, CNRS, Physicochimie des Electrolytes et Nanosyst\`emes Interfaciaux (PHENIX), 4 Place Jussieu, 75005 Paris, France}

\date{\today}

%\begin{abstract}
%  We calculate the exact time dependence of the cumulants of a tracer particle in a dense lattice gas of hardcore particles, in different geometries (one-dimensional, comb-like, and two-dimensional). More precisely, we calculate the cumulant generating function associated to the position of a tagged particle at arbitrary time, and at leading order in the density of vacancies on the lattice. Our approach gives access to the short-time dynamics of the cumulants of the tracer position -- a regime in which few results are known. Our analytical results are confronted to numerical simulations. In 1D, we consider the case of a tracer that may be biased, and we show that the full time dependence of all the cumulants takes a remarkably simple form. In other geometries, we consider a symmetric tracer particle, compute the cumulant generating function in Laplace domain, and unveil the full time-dependence of the first cumulants.
%\end{abstract}

\maketitle

\tableofcontents

\section{From a single vacancy to the dense regime}
\label{SM_singlevactodense}

\emph{In this Section, we derive Eq. (2) from the main text.}\\

Let us consider a system of finite size $N$ in which all the sites are occupied except $M$ of them (Fig. 1). We call these empty sites \emph{vacancies}, and their fraction is denoted by $\rho_0 = M/N = 1-\rho$. The high-density regime of the SEP corresponds to $\rho\to 1$. Instead of looking at the motion of the particles, one can equivalently study the motion of the vacancies. The later perform (a priori correlated) random walks on the lattice.

The tracer is initially at the origin, and its displacement at time $t$ is $\XX(t)$. This displacement can be said to be generated by the random walks of the vacancies: the tracer moves by exchanging its position with that of a neighboring vacancy. We number the vacancies and call $\XX_j(t)$ the displacement of the TP generated by the $j$-th vacancy. We have $\XX(t) = \XX_1(t) + \dots \XX_M(t)$.

The initial positions of the vacancies are called $\YY_j$. $P(\XX|\YY_1,\dots,\YY_M,t)$ is the probability of a displacement $\XX$ at time $t$ knowing the initial positions of the vacancies. Similarly, $\mathcal{P}(\XX_1,\dots,\XX_M|\YY_1,\dots,\YY_M,t)$ is the probability that up to time $t$ vacancies induced displacements $\{\XX_j\}$ of the TP knowing their initial positions
(see Fig.~1).
By definition,
\begin{equation} \label{eq:dense_linkP}
P (\XX|\{\YY_j\},t) =
\sum_{\YY_1,\dots, \YY_M} \delta_{\XX, \XX_1+\dots+\XX_M}
\mathcal{P} (\XX_1,\dots,\XX_M|\YY_1,\dots,\YY_M,t).
\end{equation}

Now, in the high density limit ($\rho_0 = M/N\to 0$), we assume that
the vacancies perform independent random walks and interact independently with the TP.
We neglect event of order $\mathcal{O}(\rho_0^2)$ in which two vacancies interact with each other, compared to events of order $\mathcal{O}(\rho_0)$ in which one vacancy interacts with the TP. This gives exact results at linear order in the density of vacancies $\rho_0$. We call $p_1(\XX|\YY,t)$ the probability that, in a system with a single vacancy initially at $\YY$, the TP has displacement $\XX$ at time $t$. Our assumption leads to
\begin{equation}
 \mathcal{P}(\XX_1,\dots,\XX_M|\YY_1,\dots,\YY_M,t) \underset{\rho_0\to 0}{\sim} \prod_{j=1}^M p_1(\XX_j|\YY_j,t)
\end{equation}
with $\rho_0 = 1-\rho$.
%Note that there are only two values of $Y$ for which $p^{(t)}_{Z}(Y)$ is non-zero ($Y = 0$ and $\pm 1$ for $Z\lessgtr 0)$). As we show in the following our problem now becomes much simpler.
Eq.~\eqref{eq:dense_linkP} now gives
\begin{equation}
P (\XX|\YY_1,\dots,\YY_M,t) \underset{\rho_0\to 0}{\sim}
\sum_{\XX_1,\dots, \XX_M} \delta_{\XX, \XX_1+\dots+\XX_M}
 \prod_{j=1}^M p_1(\XX_j|\YY_j,t).
\end{equation}
We define the Fourier transform $\tilde f(\kk) = \sum_{\XX=-\infty}^\infty \ex{\ii \kk\cdot\XX}f(\XX)$ and obtain
\begin{equation} \label{eq:1tp_link_prop}
\tilde P (\kk|\YY_1,\dots,\YY_M,t) \underset{\rho_0\to 0}{\sim}\prod_{j=1}^M \tilde p_1(\kk|\YY_j,t).
\end{equation}
We consider an initial condition in which the vacancies have equal probability to be on any site (except the origin). This corresponds to an equilibrated system and is known in the literature as \emph{annealed} initial conditions. It can be opposed to the case of an initial frozen repartition of vacancies on the lattice, usually refered to as \emph{quenched} initial conditions. Note that the choice of the type of initial conditions, annealed or quenched, can have a dramatic effect on the statistics of the position of the tracer, as studied recently in 1D geometries \cite{Krapivsky2014,Krapivsky2015,Sadhua,Poncet2021a}.

The cumulant-generating function of $\XX(t)$ is the logarithm of the average of $\tilde P(\kk|\YY_1,\dots,\YY_M,t)$,
\begin{equation}
 \psi(\kk,t) = \ln \tilde P(\kk,t),
 \end{equation}
 where
 \begin{equation}
 \tilde P(\kk,t) \equiv \frac{1}{({N}-1)^M} \sum_{\YY_1, \dots, \YY_M\neq 0} \tilde P (\kk|\YY_1,\dots,\YY_M,t).
\end{equation}
In the limit $\rho_0\to 0$, we obtain
\begin{equation}
 \tilde P(\kk,t)  \underset{\rho_0\to 0}{\sim} \left[\frac{1}{{N}-1} \sum_{\YY \neq 0} \tilde p_1(\kk|\YY,t)\right]^M = \left[1 +\frac{1}{{N}-1} \sum_{\YY\neq 0} \left(\tilde p_1(\kk|\YY,t)-1\right)\right]^M.
 \label{Ptilde_before_thermolim}
\end{equation}
We consider the large-size limit $M, {N}\to\infty$ with $\rho_0 = M/{N} = 1-\rho$ constant. We obtain an expression for the propagator $\tilde{P}(\kk,t)$ in the high-density limit:
\begin{equation}
    \tilde{P}(\kk,t) \sim \exp \left(\rho_0\sum_{\YY\neq 0} \left(\tilde p_1(\kk|\YY,t)-1\right)  \right),
\end{equation}
and for the cumulant-generating function:
\begin{equation} \label{eq:1tp_expr_psi_sup}
\lim_{\rho_0\to 0} \frac{\psi(\kk,t)}{\rho_0} = \sum_{\YY\neq 0} \left(\tilde p_1(\kk|\YY,t)-1\right),
\end{equation}
which coincides with Eq. (2) from the main text.

%\begin{figure}
%	\centering
%	\includegraphics[width=0.7\columnwidth]{systems_vacancies2.pdf}
%	\caption{For each of the considered geometries, the random walks of the particles are mirrored by the random walks of the vacancies (brown squares). The latter perform nearest-neighbor symmetric random walks walks (except on the two sites next to the TP in 1D, that we consider as defective sites because the tracer may be biased in this situation). At high density, the random  walks performed by the vacancies are assumed to be independent, which allows exact derivation at linear order in the density of vacancies $\rho_0$.}
%	\label{fig:1tp_vacs}
%\end{figure}

\section{Single-vacancy propagator}
\label{SM_singlevacprop}

\subsection{General relations}

Here, we  show how to express the single-vacancy propagator $p_1(\XX|\YY,t)$ in terms of first-passage time (FPT) densities associated to the random walks performed by the vacancies. In this section, we consider that there is only one vacancy on the lattice, initially at site $\YY$. Let $f(\zz|\YY,t)$ be the probability that the vacancy arrives at the origin for the first time at $t$, and $f^*(\zz|\ee_\nu|\YY,t)$ be the probability that the vacancy arrives at the origin for the first time at $t$, knowing that it was at site $\ee_\nu$ right before reaching the origin. For simplicity, we will use the notation $\ee_{-\mu} = -\ee_{\mu}$ ($\mu\in\{\pm 1,\dots,\pm d\}$). The propagator $p_1(\XX|\YY,t)$ can be decomposed over the first passage of the vacancy on the origin:
\begin{equation}
    \label{fpt_partition_1}
    p_1(\XX|\YY,t)= \delta_{\XX,\zz}\left(1-\int_0^t \dd \tau \; f(\zz|\YY,t) \right)+\sum_{\nu} \int_0^t \dd \tau \; p_1(\XX-\ee_\nu|-\ee_\nu,t)f^*(\zz|\ee_\nu|\YY,\tau)
\end{equation}
where the sum runs over all the directions $\nu\in\{\pm 1,\dots,\pm d\}$. One remarks that the same procedure can be applied to the total number $n$ of arrivals of the vacancy at the origin before time $t$:
\begin{align}
    p_1(\XX|\YY,t)=& \delta_{\YY,\zz}\left(1-\int_0^t \dd \tau \; f(\zz|\YY,t) \right)
    +\sum_{n=1}^\infty \int_0^\infty \dd t_1 \dots \dd t_n  \int_0^t \dd \tau \;  \delta\left(t-\sum_{i=1}^n t_i -\tau \right) \nonumber \\
    & \times \sum_{\nu_1} \dots \sum_{\nu_p} \delta_{\ee_{\nu_1}+\dots+\ee_{\nu_p},\XX} \left(1-\int_0^\tau \dd \tau'\; f(\zz|-\ee_{\nu_p},t) \right) \nonumber\\  
    &\times f^*(\zz|\ee_{\nu_p}|-\ee_{\nu_{p-1}},t_n) \dots f^*(\zz|\ee_{\nu_2}|-\ee_{\nu_1},t_2) f^*(\zz|\ee_{\nu_1}|\YY,t_2).
        \label{fpt_partition_n}
\end{align}
These equations relate the single-vacancy propagator $p_1$ to the first-passage time densities $f$ and $f^*$.

\subsection{The case of tree-like lattices}

\emph{In this Section, we derive Eq. (3) from the main text.}\\

We first consider the case of tree-like lattices. In those specific geometries, when there is one vacancy on the lattice starting from site $\YY$ ($\YY=y_1$ in 1D or $\YY=(y_1,y_2)$ on a comb), the tracer can only reach two sites: $\zz$ and $\pm \ee_1$ (depending on whether the vacancy is initially at the right or at the left of the tracer). This implies:
\begin{equation}
    f^*(\zz|\ee_\mu|\YY,t) =
    \begin{cases}
       f(\zz|\YY,t) &\text{if $\ee_\mu$ belongs to the shortest path from $\YY$ to $\zz$,} \\
       0 &\text{otherwise} .
    \end{cases}
\end{equation}
Eq. \eqref{fpt_partition_1} is then rewritten
\begin{equation}
      p_1(x|\YY,t)= \delta_{\YY,\zz}\left(1-\int_0^t \dd \tau \; f(\zz|\YY,t) \right)+ \int_0^t \dd \tau \; p_1(x-\mu|-\ee_\mu,t)f(\zz|\YY,\tau),
          \label{fpt_partition_1_tree}
\end{equation}
where $\mu \equiv \sgn(y_1)$. Using the same simplification, Eq. \eqref{fpt_partition_n} now reads
\begin{eqnarray}
p_1(y|\ee_\nu,t) &=& 
\sum_{n=0}^\infty  \delta_{y, \nu[1 - (-1)^{n+1}]} \int_0^\infty dt_1\dots dt_n  \int_0^\infty d\tau \delta\left(t - \sum_{i=1}^n t_i - \tau\right) \nonumber \\
&\times&
\left(1 - f (\zz|(-1)^n\ee_{\nu},\tau)\right) 
f (\zz|(-1)^{n-1}\ee_\nu,\tau)\dots f (\zz|-\ee_\nu,t_2) f (\zz|\ee_\nu,t_1).
\label{fpt_partition_n_tree}
\end{eqnarray}
We define the Fourier-transform in space and Laplace transform in time by
\begin{equation}
\hat{\tilde p}_1(k|\YY, u) \equiv \sum_{Y=-\infty}^\infty e^{ikx}
\int_0^\infty \dd t\;  e^{-ut} p_1(x|\YY,t).
\end{equation}
Applying it to Eqs.~ \eqref{fpt_partition_1_tree} and \eqref{fpt_partition_n_tree}, we obtain
\begin{eqnarray}
\hat{\tilde p}_1(k|\YY, u) &=& \frac{1}{u}
 + \left[\hat{\tilde p}_1(k|-\ee_\mu, u) e^{i\mu k} - \frac{1}{u}\right]  \hat f(\zz|\YY,u), \\
\hat{\tilde p}_1(k|\ee_\nu, u)&=& \frac{1}{u}
 \frac{\left[1-\hat f_\nu(u) \right] + e^{i\nu k} \hat f_\nu(u)\left[1-\hat f_{-\nu}(u) \right]}{1-\hat f_1(u)\hat f_{-1}(u)},
\end{eqnarray}
where we introduce the shorthand notation $ \hat{f}_\nu(u)=f(\zz|\ee_\nu,u)$. We combine the two equations and obtain the propagator of the displacement of the TP in term of the first passage probabilities of the vacancy,
\begin{equation}
\hat{\tilde p}_1(k|\YY, u) = \frac{1}{u}\left[
1+ \left(e^{i\mu k} -1\right) \frac{1-\hat f_{-\mu}(u)}{1-\hat f_1(u)\hat f_{-1}(u)}\hat f(\zz|\YY,u) \right],
\label{eq:1tp_expr_pZ_sup}
\end{equation}
which corresponds to Eq. (3) from the main text.

\section{FPT densities in 1D}
\label{SM_FPT_1D}

\emph{In this Section, we derive Eq. (4) from the main text.}\\

We consider a SEP with only one vacancy and a TP initially at the origin and consider the random walk performed by this unique vacancy. The vacancy is surrounded by two particles with exponential clocks with ticking probability $\chi(t) = e^{-t}$ and its Laplace transform
\begin{equation}
    \hat\chi(u) = \frac{1}{1+u}.
    \label{eq:1tp_chi}
\end{equation}

Except when it is next to the biased TP, the vacancy thus perform a symmetric Montroll-Weiss walk~\cite{Hughes1995} with a distribution of jumping times given by $\chi(t)$. When the TP is not biased, the walk becomes symmetric for all sites. We first study this situation before accounting for defective sites next to the TP.\\

\paragraph{Unbiased TP}
Let us call $f^\mathrm{UB}_y(t)$ the probability of first passage at the origin at time $t$ of a vacancy initially at $y$, assuming that the TP is not biased ($p_{\pm 1} = 1/2, s=0$).
The Montroll-Weiss walk (in continuous time) of the vacancy is linked to the associated discrete-time random walk by the formula [Ref.~\cite{Hughes1995}, Eq.~(5.46)]
\begin{equation}
\hat f^\mathrm{UB}_y(u) = \hat F_y(\hat\chi(u)),
\label{discrete_cont_equivalence}
\end{equation}
where $\hat\chi$ is given by Eq.~\eqref{eq:1tp_chi}, and $\hat F_y(\xi) = \sum_{t=0}^\infty \xi^t F_y(t)$ is the discrete Laplace transform of the probability of first passage at the origin of the discrete-time walk starting from $y$. It is known to be given [Ref.~\cite{Hughes1995}, Eq.~(3.135)] by $\hat F_y(\xi) = \alpha^{|y|}$ with $\alpha = \xi^{-1}\left(1- \sqrt{1-\xi^2}\right)$.
At the end of the day, we obtain the following expression for the first passage probability that we study:
\begin{eqnarray}
\hat f^\mathrm{UB}_y(u) &=& \alpha^{|y|}, \\
\alpha &=& 1 + u - \sqrt{u(2+u)}. \label{eq:def_alpha}
\end{eqnarray}
One notes that $\alpha$ is a solution of the equation $\alpha^2 -2(1+u)\alpha + 1 = 0$,
this leads to the non trivial relation
\begin{equation} \label{eq:rel_alpha_u}
1 + u = \frac{1+\alpha^2}{2\alpha} = \frac{1}{2}\left(\alpha + \alpha^{-1}\right).
\end{equation}
Now that we have the expression for an unbiased TP, we turn to the case of a biased TP.\\

\paragraph{Biased TP}
We consider a unique vacancy on the site $\nu = \pm 1$, next to a biased TP. Two events can
happen, either the TP jumps on site $\nu$ or the particle on site $2\nu$ jumps on site $\nu$. The first event is governed by an exponential law of rate (inverse time) $p_\nu$,
while the second is associated to an exponential clock of rate $1/2$. The motion of the vacancy is thus governed by the exponential law of rate $(p_\nu + 1/2 )$, $\chi_V(t) = (p_\nu + 1/2 )e^{-(p_\nu + 1/2)t}$. When such a jump of the vacancy occurs, there is a probability $p_\nu/(p_\nu + 1/2 )$ that it is done in the direction of the TP, and
$(1/2)/(p_\nu + 1/2 )$ that it is done in the opposite direction.

We call $f_\nu(t)$ the probability of first passage of the vacancy at the origin, knowing that it starts from site $\nu$. Either it is due to the first jump of the vacancy at time $t$, or the vacancy jumps on site $2\nu$ at time $t_0 < t$, comes back to site $\nu$ by a unbiased random walk at time $t_0+t_1$ and then arrives at the origin.
This leads us to the relation,
\begin{equation}
f_\nu(t) = p_\nu e^{-(p_\nu + 1/2)t} + \int_0^t dt_0 \frac{1}{2} e^{-(p_\nu + 1/2)t_0} \int_0^{t-t_0} dt_1
f_1^\mathrm{UB}(t_1) f_\nu(t-t_0-t_1). \label{f_nu_1d}
\end{equation}
We compute the Laplace transform of this equation and remember that $\hat f_\nu^\mathrm{UB}(u) = \alpha$ with $\alpha$ given by Eq.~\eqref{eq:def_alpha}.
Moreover, $1+u$ and $\alpha$ are linked by Eq.~\eqref{eq:rel_alpha_u}. We end up with
\begin{equation}
\hat f_\nu(u) = \frac{p_\nu}{u + p_\nu + 1/2 - \alpha/2} = 
\frac{\alpha (1+\nu s)}{1+\nu s\alpha}
\end{equation}
where $s$ is the bias. In particular, as expected, if $p_\nu = 1/2$, $\hat f_\nu(u) = \hat f_\nu^\mathrm{UB}(u) = \alpha$.\\
%\begin{equation}
%f_\nu(t) = \frac{\frac{p_\nu}{1+u}}{1-\frac{\alpha}{2(1+u)}} 
%= \frac{2p_\nu}{2+u(2-\alpha)}
%\end{equation}
%If $p_\nu = 1/2$, one checks from Eq.~\eqref{eq:rel_alpha_u} that $f_\nu(t) = \hat f_\nu^\mathrm{UB}(u) = \alpha$.

Finally, considering a vacancy starting from site $y$ and decomposing over the first visit to site $\mu=\sgn(y)$, we obtain
\begin{equation}
f_y(t) = \int_0^t dt_0 f^\mathrm{UB}_{|y|-1}(t_0) f_\mu(t-t_0),
\end{equation}
and, in Laplace domain,
\begin{equation} \label{eq:1tp_res_fZ_sup}
\hat f_y(u) = \hat f^\mathrm{UB}_{|y|-1}(u) \hat f_\mu(u) = 
\frac{1+\mu s}{1+\mu s\alpha} \alpha^{|y|},
\end{equation}
which corresponds to Eq. (4) in the main text.

\section{FPT on a comb}
\label{SM_FPT_comb}

\emph{In this Section, we derive Eq. (9) from the main text.}\\

We consider the comb lattice represented on Fig. 1(b). The backbone of the lattice is the $x_1$-axis, and to each node of the backbone a one-dimensional lattice is connected and extends infinitely in directions $\pm x_2$ (called teeth).  Bath particles jump to each neighbouring site with rate $1/4$ when they are on the backbone, and with rate $1/2$ when they are on the teeth. The tracer is constrained to move on the backbone and therefore jumps with rate $1/2$ to the right or to the left. Its position is denoted by $x(t)$ and the associated generating function is $\psi(k,t)=\ln \moy{\ex{\ii k x(t)}}$.

Starting from Eq. (2), we write
\begin{equation}
\label{psi_annealed}
\lim_{\rho\to1} \frac{\psi(k,t)}{1-\rho}  = \sum_{(y_1,y_2)\neq(0,0)} \left(\widetilde{p}_{y_1,y_2}(k,t) -1\right),
\end{equation}
where the sum runs over all the sites different from the origin (all possible starting points of the vacancies). The quantity ${p}_{y_1,y_2}(x,t)$  is the probability that the tracer reaches site $x$ at time $t$ due to its interactions with a single vacancy that starts from site $(y_1,y_2)$. Following the derivation in Section \ref{SM_singlevacprop}, the Fourier-Laplace transform of the single-vacancy propagator is related to the FPT densities $\hat{f}$ through [Eq. (3)]
\begin{equation}
\label{p_LapFourier}
\hat{\widetilde{p}}_{y_1,y_2}(k,u) = \frac{1}{u} \left[1+(\ex{\ii \mu k}-1)\frac{1-\hat{f}_{-\mu}(u)}{1-\hat{f}_{1}(u)\hat{f}_{-1}(u)}  \hat{f}(0,0|y_1,y_2;u)  \right]= \frac{1}{u} \left[1+\frac{\ex{\ii \mu k}-1}{1+\hat{f}_{1}(u)}  \hat{f}(0,0|y_1,y_2;u)  \right],
\end{equation}
where $f(0,0|y_1,y_2,t)$ is the probability for a vacancy to reach the origin for the first time at time $t$ starting from site $(y_1,y_2)$. We also define $\hat{f}_{\mu}(u)=\hat{f}(0,0|\mu,0,u)$. We used the symmetry relation $\hat{f}_{1}=\hat{f}_{-1}$ to obtain the second equality in Eq. \eqref{p_LapFourier}.

Because of the tree-like structure of the lattice, there is a single path linking site $(y_1,y_2)$ to the origin $(0,0)$. This property ensures the relation \cite{Woess2000}
\begin{equation}
    f(0,0|y_1,y_2;t) = \int_0^t \dd \tau \; f(0,0|y'_1,y'_2,\tau) f(y'_1,y'_2|y_1,y_2,t-\tau),
\end{equation}
or, in Laplace space,
\begin{equation}
    \hat f(0,0|y_1,y_2;u) =  \hat f(0,0|y'_1,y'_2,u) \hat f(y'_1,y'_2|y_1,y_2, u) .
\end{equation}
This relation holds for any site $(y'_1,y'_2)$ belonging to the bath linking $(y_1,y_2)$ and $(0,0)$. Using the path decomposition $(y_1,y_2)$ $\to$ $(y_1,\sgn(y_2))$ $\to$ $(y_1,0)$ $\to$ $(\sgn(y_1),0)$ $\to$ $(0,0)$, we write
\begin{equation}
\label{}
\hat{f}(0,0|y_1,y_2;u)=
\begin{cases}
   \hat{f}_1 (u) [\hat{f}(1,0|2,0,u)]^{|y_1|-1} \hat{f}(1,0|1,1,u) [\hat{f}(1,1|1,2,u)]^{|y_2|-1} & \text{if $y_2\neq0$}, \\
   \hat{f}_1  (u)   [\hat{f}(1,0|2,0,u)]^{|y_1|-1} & \text{if $y_2= 0$}.
\end{cases}
\label{fpt_comb}
\end{equation}
 For simplicity, we introduce the following notations:
\begin{eqnarray}
\hat{f}(1,0|2,0,u) &=& \hat{f}_{\parallel} \label{def_fparallel}\\
\hat{f}(1,0|1,1,u) &=& \hat{f}_{\perp} \label{def_fperp}
\end{eqnarray}
We note that $\hat{f}(1,1|2,1,u)$, which denotes the FPT density between two neighboring site of a tooth of the comb is nothing but the FPT density between two neighboring site of a 1D lattice, and is given by
\begin{equation}
    \hat{f}(1,1|2,1,u) = \alpha = 1+u-\sqrt{u(2+u)}
\end{equation}
Finally, using the expressions of the first-passage densities [Eq. \eqref{fpt_comb}], the single-vacancy propagator [Eq. \eqref{p_LapFourier}] and of the cumulant generating function [Eq. \eqref{psi_annealed}] yields the following expression of the latter in Laplace domain:
\begin{equation}
\label{cgf_comb}
\lim_{\rho\to1} \frac{\psi(k,u)}{1-\rho}  = \frac{1}{u}\frac{1}{1+\hat{f}_{1}(u)}  \sum_{\epsilon = \pm1} \sum_{y_2=-\infty}^\infty\ \sum_{y_1=1}^\infty (\ex{\ii \epsilon k}-1)  \hat{f}(0,0|\epsilon y_1,y_2;u) 
\end{equation}
and eventually
\begin{equation}
\label{cgf_comb2}
\lim_{\rho\to1} \frac{\psi(k,u)}{1-\rho}  =
%\frac{1}{u} \frac{1}{1-\hat{f}_{1}(u)\hat{f}_{-1}(u)}\frac{1}{1-\hat{f}_\parallel(u)} \left(1+\frac{2\hat{f}_\perp(u)}{1-\alpha(u)}\right) \sum_{\mu=\pm1}(\ex{\ii\mu k}-1) \hat{f}_{\mu}(u)(1-\hat{f}_{-\mu}(u)).
\frac{1}{u} \frac{1}{1-\hat{f}_\parallel(u)} \left(1+\frac{2\hat{f}_\perp(u)}{1-\alpha(u)}\right) \frac{\hat{f}_1(u)}{1+\hat{f}_1(u)}\cdot 2(\cos k -1).
\end{equation}
The last step of the calculation is to compute the quantities $\hat{f}_{\pm 1}$, $\hat{f}_{\parallel}$ and $\hat{f}_{\perp}$:

\begin{itemize}

\item {Calculation of $\hat{f}_{ 1}$}

Here, we follow the arguments that lead to the derivation of Eq. \eqref{f_nu_1d}. Considering a vacancy initially located at site $(\pm 1,0)$ (on the backbone and right next to the tracer) and partitioning over the first jump performed by the vacancy which can either be directed towards the tracer (with rate $1/2 $) on the backbone and in the direction opposite to that of the tracer (with rate $1/4$), or sideways on the tooth of the comb (with rate $2\times1/2$), one writes
\begin{eqnarray}
 f_1(t) &= &\frac{1}{2} \ex{-7t/4} + \int_0^t \dd t_0\; \frac{1}{4}\ex{-7t/4} \int_0^{t-t_0}\dd t_1 \;f_\mu(t-t_0-t_1) f^\text{UB}_\parallel(t_1) \nonumber\\
 &&+2\int_0^t \dd t_0 \;\frac{1}{2} \ex{-7t/4} \int_0^{t-t_0}\dd t_1 \;  f_\mu(t-t_0-t_1)f^\text{UB}_\parallel(t_1)
\end{eqnarray}
where $f_\parallel$ is defined in Eq. \eqref{def_fparallel}. Taking the Laplace transform of this equation, one gets
\begin{equation}
    \hat{f}_1 (u)= \frac{1/2}{u+\frac{7}{4}-\frac{1}{4}\hat{f}_\parallel - \hat{f}_\perp}.
\end{equation}

\item {Calculation of $\hat{f}_{\parallel}$}

In order to calculate $\hat{f}_{\parallel} = \hat{f}(1,0|2,0,u)$, we consider a vacancy starting from site $(2,0)$ and, partitioning over the first jump performed by the vacancy which can either on the backbone (with rate $2\times 1/4$) or sideways on the tooth of the comb (with rate $2\times1/2$), we get
\begin{eqnarray}
 f_\parallel(t) &= &\frac{1}{4} \ex{-3t/2} + \int_0^t \dd t_0\; \frac{1}{4}\ex{-3t/2} \int_0^{t-t_0}\dd t_1 \; f_\parallel(t-t_0-t_1) f_\parallel(t_1) \nonumber\\
 &&+2\int_0^t \dd t_0 \; \frac{1}{2} \ex{-3t/2} \int_0^{t-t_0}\dd t_1 \; f_\parallel(t-t_0-t_1)f_\perp(t_1).
\end{eqnarray}
In Laplace space, one gets the equation satisfied by $\hat{f}_\parallel(u)$:
\begin{equation}
    \hat{f}_\parallel(u)^2-4\left[u+\frac{3}{2}-\hat{f}_\perp(u) \right] \hat{f}_\parallel(u) +1 =0.
\end{equation}
Choosing the solution satisfying the short-time condition $\lim_{u\to\infty} \hat{f}_\parallel(u)=0$, we get
\begin{equation}
    \hat{f}_\parallel (u) = 2 \left(u+\frac{3}{2} - \hat{f}_\perp (u) \right)  -\sqrt{4\left(u+\frac{3}{2} - \hat{f}_\perp (u) \right)^2-1}
\end{equation}

\item {Calculation of $\hat{f}_{\perp}$}

Finally, in order to calculate $\hat{f}_{\perp} = \hat{f}(1,0|1,1,u)$, we consider a vacancy starting from site $(1,1)$ and, partitioning over the first jump performed by the vacancy which can either be away from the backbone (with rate $1/2$) or to the backbone (with rate $1/4$), we get
\begin{equation}
 f_\perp(t) = \frac{1}{4} \ex{-3t/4} + \int_0^t \dd t_0\; \frac{1}{2}\ex{-3t/4} \int_0^{t-t_0}\dd t_1 \; f_\perp(t-t_0-t_1) f_\text{1D}(t_1),
\end{equation}
where $f_\text{1D}$ is the FPT density of a vacancy between two neighboring sites of a one-dimensional lattice (its Laplace transform is denoted by $\alpha(u)$ in the main text). In Laplace space, we get
\begin{equation}
    \hat{f}_{\perp}=\frac{1}{4u+3-2\alpha(u)}=\frac{\alpha}{2-\alpha}
\end{equation}

\end{itemize}

Eq. \eqref{cgf_comb2}, together with the expressions of $\hat f_\parallel$, $\hat f_\parallel$, and $\hat f_1$ above, leads after some algebra to Eq. (9) of the main text.

\section{Looped lattices :  expression of the CGF in terms of the conditional FPT $f^*$}
\label{SM_looped1}

\emph{In this Section, we derive Eq. (10) from the main text.}\\

We now turn to the case of $d$-dimensional lattices. We start from Eq. \eqref{Ptilde_before_thermolim}, which relates the propagator of the tracer position to the single-vacancy propagators in 
Fourier space $\tilde{p}_1(\kk|\YY,t)$:
\begin{equation}
 \tilde P(\kk,t) \simeq \left[1 +\frac{1}{{N}-1} \sum_{\YY\neq 0} \left(\tilde p_1(\kk|\YY,t)-1\right)\right]^M.
 \label{Ptilde_before_thermolim2}
\end{equation}
Using the Fourier transform of Eq. \eqref{}, which relate the single-vacancy propagators of the random walk of a tracer starting from an arbitrary point $\YY$ and from a site located at the vicinity of the tracer $\ee_\nu$, we find the relation 
\begin{equation}
    \tilde{p}_1(\kk|\YY,t)=1-\int_0^t\dd\tau\; f(\zz|\YY,t) + \sum_\nu \int_0^t \dd\tau\; f^*(\zz|\ee_\nu|\YY;\tau)\ex{\ii q_\nu} \tilde{p}_1(\kk|-\ee_\nu;t-\tau).
\end{equation}
Replacing $\tilde{p}_1(\kk|\YY,t)$ in Eq. \eqref{Ptilde_before_thermolim2} by this expression, and using the relation $f(\zz|\YY;t) = \sum_\nu f^*(\zz|\ee_\nu|\YY;t)$, one gets
\begin{equation}
 \tilde P(\kk,t) = \left[1 -\frac{1}{{N}-1} \sum_\nu \int_0^t\dd\tau \; [1-\ex{\ii q_\nu} \tilde{p}_1(\kk\-\ee_\nu;t-\tau)] \sum_{\YY\neq\zz} f^*(\zz|\ee_\nu|\YY;\tau)   \right]^M.
\end{equation}
Using the equivalence between directions $\pm \ee_\nu$, defining (for $j=1,\dots,d$)
\begin{equation}
    \Delta(\kk|\ee_j,t) = 2-\ex{\ii q_j}\tilde{p}_1(\kk|-\ee_j;t)-\ex{-\ii q_j}\tilde{p}_1(\kk|\ee_j;t),
    \label{Delta_def}
\end{equation}
and taking the thermodynamics limit ($M,N\to\infty$ with fixed $\rho=M/N$), one gets
\begin{equation}
 \tilde P(\kk,t) = \exp \left[ - \rho\sum_{j=1}^d \int_0^t\dd\tau \; \Delta(\kk|\ee_j,t-\tau) f'_\nu(\tau)   \right].
\end{equation}
where we defined
\begin{equation}
 f'_\nu(t) = \sum_{\YY\neq\zz} f^*(\zz|\ee_\nu|\YY;t).
 \label{def_fp}
\end{equation}
In Laplace domain, the CGF then reads:
\begin{equation}
 \lim_{\rho_0 \to 0} \frac{\hat{\psi}(\kk,u)}{\rho_0} = -\sum_{j=1}^d \hat{\Delta}(\kk|\ee_j,u) \hat{f}'_j(u). 
    \label{cgf_Delta_fp_sup}
\end{equation}
The last step of the calculation consists in expressing  $\hat{\Delta}(\kk|\ee_j,u)$ in terms of the conditional FPT densities $f^*(\zz|\ee_\nu|\ee_\mu,t)$. Taking the Fourier-Laplace transform of Eq. \eqref{fpt_partition_n},  we get
\begin{equation}
\label{eq:P1def}
\hat{\tilde{p}}_1(\kk|\YY;u)=\frac{1}{u}\left(1+\sum_{\mu}V_{ \mu}(\kk;u)f^*({\bf
0}|\ee_\mu|\YY;u)\right).
\end{equation}
where we defined
\begin{equation}
\label{eq:Udef}
V_{\mu}(\kk;u)\equiv \sum_{ \nu} 
\left[1-\ex{-\ii \kk \cdot \ee_\nu}\right]\left\{[{\bf I} -{\bf T}({\kk};u)]^{-1}\right\}_{\nu,\mu}  \ex{\ii {\kk} \cdot \ee_\mu}.
\end{equation}
$\mathbf{I}$ is the identity of size $2d$ and the matrix ${\rm {\bf T}}(\kk;u)$ has the entries $\left[{\rm {\bf T}}(\kk;u)\right]_{\mu,\nu}$  defined by
\begin{equation}
\label{Tdef}
\left[{\rm {\bf T}}(\kk;u)\right]_{\nu,\mu} = \ex{\ii \kk \cdot {\bf
e_{\nu}}}  \hat{f}^*({\bf 0}|\ee_\nu|-\ee_\mu ; u) = \int_0^\infty \dd t \; 
f^*_t({\bf 0}|\ee_\nu |-\ee_\mu,t) \ex{-ut}.
\end{equation}
Taking the Laplace transform of Eq. \eqref{Delta_def}, and using the expression of $\hat{\tilde{p}}_1$ [Eq. \eqref{eq:P1def}] yields
\begin{equation}
    \hat{\Delta}(\kk|\ee_j,u)=\frac{2(1-\cos q_j)}{u}-\frac{1}{u} \sum_{\mu} V_\mu (\kk,u) [\ex{\ii q_j} \hat{f}^*(\zz|\ee_\mu|-\ee_j,u)+\ex{-\ii q_j} \hat{f}^*(\zz|\ee_\mu|\ee_j,u)]
    \label{Delta_fstar}
\end{equation}.
We then have an expression of the CGF in terms of the conditional first-passage densities $f^*$ (Eq. (10) from the main text).

\section{Conditional FPT densities $f^*$ on a looped lattice}
\label{SM_looped2}

\emph{In this Section, we derive Eqs. (13) and (14) from the main text.}\\

%The final step of the calculation consists in determining the conditional FPT $f^*$ in terms of well-known quantities, namely the propagators associated to a discrete-time random walk on a lattice. The following relation
%\begin{equation}
%    \int_0^t \dd t_0 \; \frac{1}{2d} \chi(t_0) p(\ee_\mu|\yy,t-t_0) = \int_0^t \dd t_0 f^*(\zz|\ee_\mu|\yy,t_0)\Psi(t-t_0)+\int_0^t\dd t_0 \int_0^{t-t_0}\dd t_1\; \frac{1}{2d} \chi(t-t_0-t_1) f(\zz|\yy,t_0) p(\ee_\mu|\zz,t_1),
%    \label{partition_fstar}
%\end{equation}
%where $\Psi(t) = 1-\int_0^t \dd t'\; \chi(t')$ is the probability that the walker did not move during a time $t$. Both the lhs and the rhs of Eq. \eqref{partition_fstar} are equal to the probability that a walker, starting from $\yy\neq \zz$ at $t=0$, reaches site $\zz$ at $t$ and was at site $\ee_\mu$ right before its last jump. The rhs is obtained by partitioning over the first visits of the walker to the origin. 
Taking the Laplace transform of this Eq. (12) in the main text, using the renewal equation $\hat{f}(\yy|\zz,u)=\hat{p}(\yy|\zz,u)/\hat{p}(\zz|\zz,u)$ (valid for $\yy\neq\zz$ \cite{Hughes1995}), and using $\hat{\Psi}(u) = (1-\hat{\chi}(u))/u$, we get
\begin{equation}
  \hat{f}^*(\zz|\ee_\mu|\yy,u) = \frac{1}{2d} \frac{u\hat\chi(u)}{1-\hat\chi(u)} \left[\hat{p}(\ee_\mu-\yy|\zz,u)-\frac{\hat{p}(\yy|\zz,u)\hat{p}(\ee_\mu|\zz,u)}{\hat{p}(\zz|\zz,u)}  \right]
\end{equation}
where $p(\rr|\rr_0;t)$ is the propagator associated to a simple random walk on the considered lattice (probability to find a walker at site $\rr$ at time $t$ knowing that it started from site $\rr_0$ at time $t=0$), and we used the translational invariance of the lattice (i.e. $p(\rr|\rr_0;t)=p(\rr-\rr_0|\zz;t)$ for any two sites $\rr$ and $\rr_0$).

The Laplace transform of the continuous-time propagator can be related to the generating function $\hat{P}(\rr|\rr_0;\xi) = \sum_{n=0}^\infty P_n(\rr|\rr_0)\xi^n$ associated to the discrete-time propagator $P_n(\rr|\rr_0)$ (probability to find the walker at site $\rr$ after $n$ steps knowing that it started from site $\rr_0$) through the relation \cite{Hughes1995}:
\begin{equation}
    \hat{p}(\rr|\rr_0,u)=\frac{1-\hat{\chi}(u)}{u}\hat{P}(\rr|\rr_0;\hat{\chi}(u)),
\end{equation}
which yields 
\begin{equation}
  \hat{f}^*(\zz|\ee_\mu|\yy,u) =\frac{\hat{\chi}(u)}{2d}  \left[  \hat{P}(\ee_\mu-\yy|\zz;\hat{\chi}(u))-\frac{\hat{P}(\yy|\zz;\hat{\chi}(u))\hat{P}(\ee_\mu|\zz;\hat{\chi}(u))}{\hat{P}(\zz|\zz;\hat{\chi}(u))} \right].,
%  \label{fhat_Phat}
\end{equation}
which coincides with Eq. (13) in the main text.

We finally compute $\hat{f}'_\mu(u)$, defined in Eq. \eqref{def_fp}. To this end, we must use the normalisation condition $\sum_{\rr}P_n(\rr|\rr_0)=1$, which reads, in terms of the generating function associated to $P_n$:
\begin{equation}
    \sum_{\rr}\hat{P}(\rr|\rr_0,\xi) = \frac{1}{1-\xi}.
\end{equation}
With Eq. (13), we get the simple expression
\begin{equation}
    \hat{f}'_\mu(u) = \frac{1}{2d} \frac{\hat{\chi}(u)}{1-\hat{\chi}(u)} \left[1- \frac{\hat{P}(\ee_\mu|\zz,\hat{\chi}(u))}{\hat{P}(\zz|\zz,\hat{\chi}(u))} \right].
   %   \label{fprimehat_Phat}
\end{equation}
which is Eq. (14) in the main text.

\section{Application to a 2D lattice}
\label{SM_2D}

\emph{In this Section, we give the expression of the CGF of the tracer particle on a 2D lattice.}\\

We now consider the example of the infinite 2D lattice. Making use of the symmetries of this lattice, one can show that the matrix $\mathbf{T}(\kk;u)$ (defined in Eq. \eqref{Tdef}) takes the simple form
\begin{equation}
    \mathbf{T}(\kk;u) = \begin{pmatrix}
    \ex{\ii k_1}a(u)  & \ex{\ii k_1}a(u)  & \ex{\ii k_1}c(u)  & \ex{\ii k_1}c(u)     \\
    \ex{-\ii k_1}a(u)  & \ex{-\ii k_1}a(u)  & \ex{-\ii k_1}c(u)  & \ex{-\ii k_1}c(u)     \\
    \ex{\ii k_2}c(u)  & \ex{\ii k_2}c(u)  & \ex{\ii k_2}b(u)  & \ex{\ii k_2}a(u)     \\
    \ex{-\ii k_2}c(u)  & \ex{-\ii k_2}c(u)  & \ex{-\ii k_2}a(u)  & \ex{-\ii k_2}b(u)     \\
\end{pmatrix}
\end{equation}
where we introduce shorthand notations for the following conditional FPT densities, which are determined in terms of the generating functions $\hat{P}$ using Eq. (13):
\begin{eqnarray}
\label{}
   a(u)  &=&f^*(\zz|\ee_1|\ee_1,u)=\frac{\hat{\chi}(u)}{4}\left[  \hat{P}(\zz|\zz;\hat{\chi}(u))-\frac{\hat{P}(\ee_1|\zz;\hat{\chi}(u))^2}{\hat{P}(\zz|\zz;\hat{\chi}(u))} \right],  \\
   b(u)  &=&f^*(\zz|\ee_1|-\ee_1,u)=\frac{\hat{\chi}(u)}{4} \left[  \hat{P}(2\ee_1|\zz;\hat{\chi}(u))-\frac{\hat{P}(\ee_1|\zz;\hat{\chi}(u))^2}{\hat{P}(\zz|\zz;\hat{\chi}(u))} \right],   \\
   c(u)  &=&f^*(\zz|\ee_1|\ee_2,u)=\frac{\hat{\chi}(u)}{4}\left[  \hat{P}(\ee_1+\ee_2|\zz;\hat{\chi}(u))-\frac{\hat{P}(\ee_1|\zz;\hat{\chi}(u))^2}{\hat{P}(\zz|\zz;\hat{\chi}(u))} \right]  .
\end{eqnarray}
Since we are only interested in the cumulants of the position projected onto one direction of the lattice, we only consider the dependence of the CGF on the component $k_1$ and set $k_2=0$ for simplicity. Using Eq. (10) (together with Eqs. \eqref{def_fp} and \eqref{Delta_fstar} yields the following expression of the CGF in terms of the conditional FPT densities:
\begin{equation}
    \lim_{\rho_0 \to 0} \frac{\hat{\psi}(k_1,k_2=0,u)}{\rho_0} = 
    \frac{1}{2u} \frac{\hat\chi(u)}{1-\hat\chi(u)} \frac{(1-\cos k_1)(a-b-1)[(a+b-1)^2-4c^2]}{[2b(a+b-1)-4c^2]\cos k_1+(a+b-1)(a^2-b^2-1)-4(a-b)c^2}.
    %\label{cgf_Delta_fp}
\end{equation}
Finally, the CGF is only expressed in terms of 4 distinct propagators: $\hat{P}(\zz|\zz;\xi)$, $\hat{P}(\ee_1|\zz;\xi)$, $\hat{P}(2\ee_1|\zz;\xi)$ and $\hat{P}(\ee_1+\ee_2|\zz;\xi)$. There exists relations between these propagators \cite{Brummelhuis1988,Brummelhuis1989a}, as well as explicit expressions of them in terms of integrals, that eventually allow a fully explicit determination of the CGF.

\section{Explicit expression of $g(\xi)$ in terms of elliptic integrals}
\label{SM_explicit_g}

The function $g$, defined in Eq. (16) of the main text as
\begin{equation}
    g(\xi) \equiv \frac{1}{2}[\hat{P}(\zz|\zz;\xi)-\hat{P}(2\ee_1|\zz;\xi)]= \frac{1}{(2\pi)^2} \int_{-\pi}^\pi \dd q_1 \int_{-\pi}^\pi \dd q_2 \; \frac{\sin^2 q_1}{1- \frac{\xi}{2}(\cos q_1 + \cos q_2)}.
   % \label{def_g}
\end{equation}
 can be expressed in terms of elliptic integrals:
\begin{equation}
    g(\xi) = \frac{4}{\xi^2} + \frac{4}{\pi\xi^2\sqrt{1-\xi^2}}\left[(1-\xi) \text{K}\left( \frac{\ii \xi}{\sqrt{1-\xi^2}} \right) -(1-\xi^2)\text{E}\left( \frac{\ii \xi}{\sqrt{1-\xi^2}}\right)-2\Pi\left(\frac{\xi}{\xi-1},\frac{\ii \xi}{\sqrt{1-\xi^2}} \right)\right],
\end{equation}
where we use the following expressions for the elliptic integrals:
\begin{equation}
    \text{K}(k)= \int_0^1 \frac{\dd t}{\sqrt{1-t^2}\sqrt{1-k^2 t^2}} 
    \qquad ; \qquad 
    \text{E}(k)= \int_0^1 \dd t \frac{\sqrt{1-k^2 t^2}}{\sqrt{1-t^2}} 
    \qquad ; \qquad 
    \Pi(\nu,k)= \int_0^1 \frac{\dd t}{(1-\nu t^2)\sqrt{1-t^2}\sqrt{1-k^2 t^2}}.
\end{equation}
The asymptotic expansions of $g(\xi)$ read 
\begin{eqnarray}
    g(\xi)&\underset{\xi\to1}{=} &\left( 2- \frac{4}{\pi}\right)+ \frac{2}{\pi} (1-\xi) \ln (1-\xi) + \mathcal{O}(1-\xi), \\
    g(\xi)& \underset{\xi\to0}{=} &\frac{1}{2}+\frac{3}{32} \xi^2+ \mathcal{O}\left( \xi^3 \right).
\end{eqnarray}

\section{Numerical simulations}
\label{SM_simus}

%\comment{Add details about the numerical simulations}

\begin{figure}
	\centering
	\includegraphics[width=0.8\columnwidth]{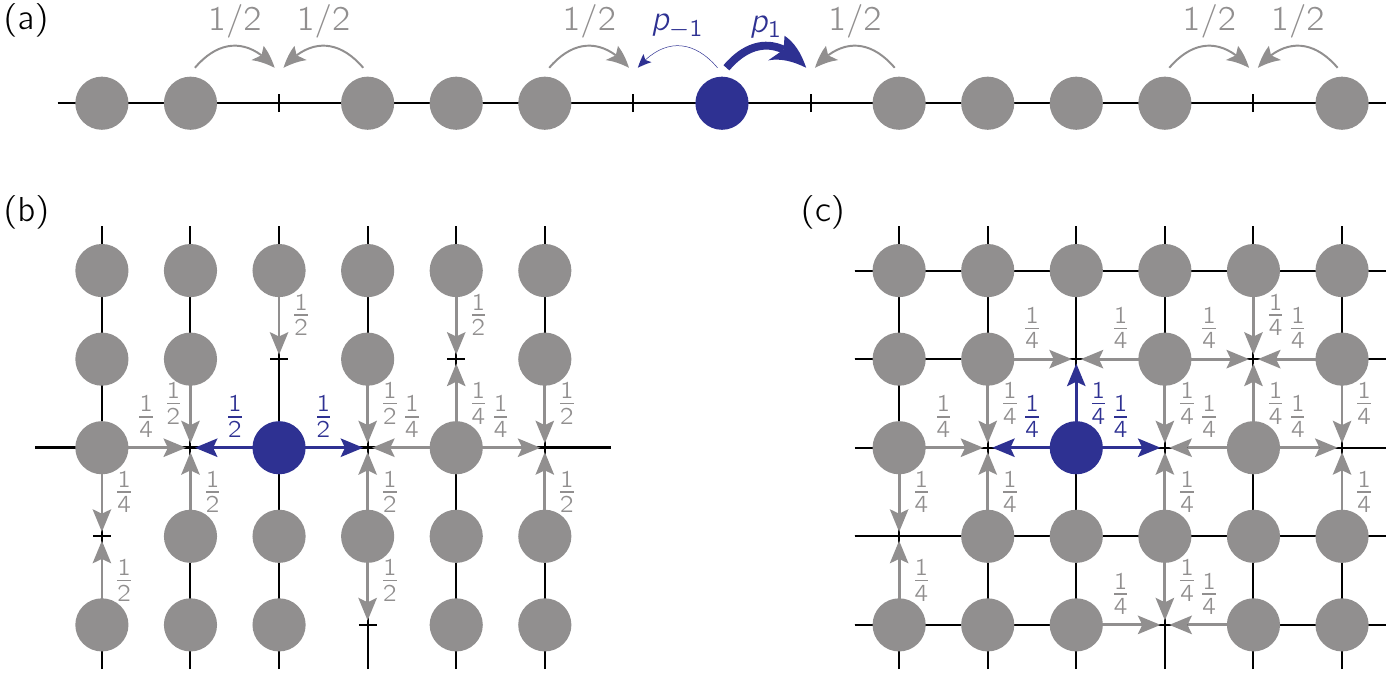}
	\caption{(a) Symmetric Exclusion Process (SEP) with a biased TP. The bath particles jump on neighboring site with rate $1/2$, whereas the jump rates of the TP are $p_{\pm 1} = (1\pm s)/2$ where $s$ is the bias. (b) Tracer diffusing on a crowded comb lattice. Particles jump on neighboring sites with rate $1/4$ (resp. $1/2$) when they are on the backbone (resp. the teeth) of the comb. The tracer is constrained to move on the backbone of the lattice. (c) Tracer diffusion on a crowded 2D lattice. The bath particles and the tracer jump on each neighboring site with rate $1/4$.}
	\label{fig:1tp_fig}
\end{figure}

The simulations of the SEP are performed on a periodic ring of size $N$, with $M=\rho N$ particles at average density $\rho$. In Fig.~2, $N=2000$ and $M=1960$ ($\rho_0 = 0.02$). The particles are initially placed uniformly at random. The jumps of the particles are implemented as follow. One first picks a particle uniformly at random. Then a direction (left or right) is chosen according to probabilities $1/2$ and $1/2$ for bath particles; $p_1$ and $p_{-1}=1-p_1$ for the tracer. If the chosen particle has no neighbor in that direction, the jump is performed, otherwise it is rejected. In both cases, the time of the simulation is incremented by a random number picked from an exponential distribution of rate $N$.
We keep track of the particle initially at the origin (the tracer) and compute the moments of its displacement, averaging over $2\cdot 10^6$ simulations.

The simulations of the comb and of the 2D lattice (Fig.~3) are performed in a similar way. Starting from a uniform random configuration, particles are chosen uniformly at random and try to jump to neighboring sites with probabilities described on Fig.~\ref{fig:1tp_fig}. Hard-core exclusion is enforced. In both cases (comb and 2D lattice), we use a periodic grid of size $100\times 100$ with $9900$ particles ($\rho_0 = 0.01$) and average over $4\cdot 10^6$ simulations.

% \section{2D lattice: single-vacancy propagator}

% \section{Asymptotic expansion of $g_{00}$ and $g_{20}$}

% We give below the asymptotic expansions of $g_{00}$ and $g_{20}$ in Laplace space ($u\to0$ and $u\to\infty$)
% \begin{eqnarray}
% \label{ }
% g_{00}(u) &=& \int_{-\pi}^\pi \frac{\dd q_1}{2\pi}\int_{-\pi}^\pi \frac{\dd q_2}{2\pi}\frac{1}{1-\frac{1}{2(1+u)}(\cos q_1+\cos q_2)}=\frac{2}{\pi} \mathrm{K}\left( \frac{1}{(1+u)^2}\right), \\
% &\underset{u\to0}{=}&\frac{1}{\pi}\ln \frac{8}{u}-\frac{1}{2\pi}u\ln u + \mathcal{O}(u)\\
% &\underset{u\to\infty}{=}& 1+\frac{1}{4u^2}+ \mathcal{O}\left( \frac{1}{u^3} \right)
% \end{eqnarray}

% \begin{eqnarray}
% \label{ }
% g_{20}(u) &=& \int_{-\pi}^\pi \frac{\dd q_1}{2\pi}\int_{-\pi}^\pi \frac{\dd q_2}{2\pi}\frac{\sin^2 q_1}{1-\frac{1}{2(1+u)}(\cos q_1+\cos q_2)}, \\
% &\underset{u\to0}{=}& \left( 2- \frac{4}{\pi}\right)+ \frac{1}{2\pi} u \ln u + \mathcal{O}(u)\\
% &\underset{u\to\infty}{=}&\frac{1}{2}+\frac{3}{32 u^2}+ \mathcal{O}\left( \frac{1}{u^3} \right)
% \end{eqnarray}

%\section*{References}

\bibliographystyle{apsrev4-1}
\bibliography{minimal_bib.bib}